%% file: PSVM-ECC-final.tex
\documentclass[letterpaper, 10 pt, conference]{ieeeconf}  

\IEEEoverridecommandlockouts                              
\usepackage{lmodern}
\usepackage[T1]{fontenc}
\usepackage{graphicx}
\usepackage{epsfig} 
\usepackage{fancyhdr}
\usepackage{mathptmx} 
\usepackage{times} 
\usepackage{amsmath} 
\usepackage{amssymb}  
\usepackage{mathrsfs}
\usepackage{url}
\usepackage{bm}
\usepackage{theorem}
\usepackage{booktabs} 
\usepackage{xcolor}
\usepackage{epstopdf}
\usepackage{algorithm}
\usepackage{algorithmic}

\usepackage{subfig}

\input{mymacros_2cln}

\input{lpsymbols}

\makeatletter

\newcommand{\Rmnum}[1]{\expandafter\@slowromancap\romannumeral #1@}
\makeatother

\title{\LARGE \bf Prescriptive Cluster-Dependent Support Vector Machines with an
  Application to Reducing Hospital Readmissions~\authorrefmark{1} \thanks{* Research
    partially supported by the NSF under grants DMS-1664644, CNS-1645681,
    IIS-1237022, and CCF-1527292, by the ONR under MURI grant N00014-16-1-2832, by
    the NIH under grant 1UL1TR001430 to the Clinical \& Translational Science
    Institute at Boston University, and by the Boston University Digital Health
    Initiative.}  }

\author{Taiyao Wang and
  Ioannis Ch. Paschalidis,~\IEEEmembership{Fellow,~IEEE}
  \thanks{Department of Electrical
      and Computer Engineering, Division of Systems Engineering, and
      Center for Information and Systems Engineering, 
      Boston University, Boston, MA 02215, USA. E-mail: {\tt \{wty,
      yannisp\}@bu.edu, url: 
      {\tt http://sites.bu.edu/paschalidis/}.}}
      }

\begin{document}

\maketitle
\thispagestyle{empty}
\pagestyle{empty}

{\allowdisplaybreaks

\begin{abstract}
We augment linear Support Vector Machine (SVM) classifiers by adding three
important features: (i) we introduce a regularization constraint to induce a
sparse classifier; (ii) we devise a method that partitions the positive class
into clusters and selects a sparse SVM classifier for each cluster; and (iii)
we develop a method to optimize the values of controllable variables in order
to reduce the number of data points which are predicted to have an undesirable
outcome, which, in our setting, coincides with being in the positive
class. The latter feature leads to personalized prescriptions/recommendations.
We apply our methods to the problem of predicting and preventing hospital
readmissions within 30-days from discharge for patients that underwent a
general surgical procedure. To that end, we leverage a large dataset
containing over 2.28 million patients who had surgeries in the period
2011--2014 in the U.S. The dataset has been collected as part of the American
College of Surgeons National Surgical Quality Improvement Program (NSQIP).
\end{abstract}


\section{Introduction}

The {\em Support Vector Machine (SVM)}~\cite{cortes1995support} is a binary
classifier, widely used in practice due to its tractability for large scale
problems. To obtain an SVM classifier, one needs to solve a convex quadratic problem
with linear constraints, which can be done for large problem instances involving
thousands of samples and hundreds of variables for each sample.  The goal of this
paper is to exploit the SVM framework to move beyond predictions and attempt to
``control'' future outcomes by appropriately modifying some of the key predictive
variables. To that end, we develop a new method we call {\em Prescriptive Support
  Vector Machine (PSVM)}.

We will apply the new method to an important problem in health care; preventing
hospital readmissions. The need for systematic, quantitative methods for addressing
health care problems is compelling. An estimated \$3 trillion is spent annually on
health care in the U.S., a value that exceeds 17\% of the U.S. Gross Domestic Product
(GDP) -- by far the largest among the 13 high-income Organization for Economic
Cooperation and Development countries. The Centers for Medicare and Medicaid Services
have identified hospital readmissions, defined as an additional admission
to address the same issue within 30 days after discharge, as an important and
potentially preventable source of excessive resource utilization and increased cost
of care~\cite{Readmissions-Reduction}.

An analysis of 2005 Medicare claims demonstrated that about 75\% of 30-day
readmissions, representing about \$12 billion in Medicare spending, were potentially
preventable~\cite{james2013medicare}. As a result, through the enactment of the
Readmissions Reduction Program section of the Affordable Care Act of 2012,
readmissions have been increasingly used as a quality of care metric, and their
reduction is mandated for certain diseases~\cite{Readmissions-Reduction}. In this
context, many surgical departments in the U.S. are establishing processes aimed at
reducing 30-day readmissions. We refer to~\cite{pas-hbr-17} for a general discussion
of the benefits and some potential risks associated with the application of health
analytics.

Several works exploit classical machine learning approaches, such as random forests,
gradient tree boosting, logistic regression, linear and kernelized SVM, and related
methods for predicting 30-day readmissions in patients with heart failure
\cite{amarasingham2010automated,frizzell2017prediction,golas2018machine}.  Recently,
the authors developed an interpretable classification approach to predict chronic
disease hospitalizations based on past Electronic Health Records (EHRs), establishing
convergence, sample complexity and generalization guarantees
\cite{ACC-2018-PIEEE,brisimi2018predicting,xu2016joint}.  Interpretability is indeed critical for medical
and health informatics, as well as other areas, e.g., safety and security management.
Without interpretable models, physicians may not use ``black box'' predictions even
if they are highly accurate.

In this paper, we augment earlier SVM-based predictive analytics along three
directions. First, we use a sparsity-inducing $\ell_1$-norm-based constraint to
obtain sparse classifiers which can generalize better out-of-sample and provide
interpretability. Second, we leverage our work in \cite{ACC-2018-PIEEE,xu2016joint}
to solve a joint clustering and classification problem and discover hidden clusters
in the positive class and corresponding, per-cluster SVM-based classifiers. The third
direction is the development of {\em prescriptive analytics}. In our setting, this
consists of a method which leverages the SVM-based predictive model to devise
personalized interventions with the potential to prevent a readmission by
controlling/optimizing the value of some variables characterizing the patient.

There have only been very few works focusing on so-called {\em prescriptive
  analytics}. An example is \cite{bertsimas2014predictive,bertsimas2017bootstrap},
which develop a data-driven framework to prescribe an optimal decision in a setting
where the cost depends on uncertain problem parameters that need to be learned from
data.

We apply our methods to a data set containing over 2.28 million patients who had
surgeries in 2011--2014 in the U.S. The data are collected as part of American
College of Surgeons (ACS) National Surgical Quality Improvement Program
(NSQIP)~\cite{ingraham2010quality}. Earlier work studied risk factors for 30-day
readmissions for categories of surgical patients, e.g., orthopaedic trauma
injuries~\cite{sathiyakumar2015asa}, knee and hip
arthroplasty~\cite{pugely2013incidence}, and ventral hernia
repair~\cite{lovecchio2014risk}. A simple readmission score using few variables was
developed based on 2011 NSQIP data only in~\cite{lucas2013assessing}.  To the best of
our knowledge, our work is the first to develop analytics for 30-day readmissions
after general surgery using millions of NSQIP records.

The remainder of this paper is organized as follows. Sec.~\ref{sec:pred} reviews
SVM-based classification and the joint clustering and classification
method~\cite{xu2016joint}. These are key building blocks for the prescriptive method
which is presented in Sec.~\ref{sec:pres}. The data and pre-processing steps
are outlined in Sec.~\ref{sec:data}. Experimental results are in
Sec.~\ref{sec:num} and conclusions in Sec.~\ref{sec:con}.

\textbf{Notation:} All vectors are column vectors and are denoted by bold lowercase
letters.  For economy of space, we write $\bx = (x_1, \ldots, x_{\text{dim}(\bx)})$
to denote the column vector $\bx$, where $\text{dim}(\bx)$ is the dimension of $\bx$.
We use prime to denote the transpose of a vector.  Unless otherwise specified,
$\|\cdot\|$ denotes the $\ell_2$ norm and $\|\cdot\|_1$ the $\ell_1$ norm.  We will
use $\|\bx\|_p=(\sum_{i=1}^{\text{dim}(\bx)} |x_i|^p)^{1/p}$ to denote the $\ell_p$
norm, where $p\geq 1$. We will also use the notation $[N]$ for the set
$\{1,\ldots,N\}$.


\section{SVM based predictive analytics} \label{sec:pred}

The SVM algorithm \cite{cortes1995support} seeks a separating hyperplane
in the variable space, so that data samples from the two different
classes reside on two different sides of the hyperplane. The minimum
over all the distances from the input data samples to the hyperplane is
called {\em margin}. The goal of SVM is to find the optimal hyperplane
that has the maximum margin.  In cases where data samples are
neither linearly nor perfectly separable, the {\em soft-margin} SVM
tolerates misclassification errors and can leverage kernel functions to
map the features into a higher dimensional space where linear
separability is possible ({\em kernelized
  SVMs})~\cite{cortes1995support}.

\subsection{SLSVM: Sparse Linear SVM} \label{sec:slsvm}

Following \cite{ACC-2018-PIEEE,xu2016joint} and our interest in interpretable
classifiers, we formulate a Sparse version of Linear SVM (SLSVM) as follows. We are
given training data $\bx_i\in\mathbb{R}^D$ and labels $y_i\in\{-1,1\}$, $i=1,\dots,
n$, where $\bx_i$ is the vector of variables characterizing the $i$th patient and
$y_i=1$ (resp., $y_i=-1$) indicates that the patient is (resp., is not)
readmitted. We will refer to the class with labels equal to $1$ as the positive class
and the other class as the negative class.

We seek to find a hyperplane orthogonal to some vector $\bbeta \in
\mathbb{R}^D$ that passes from $-\beta_0 \in \mbb{R}$, which can be done
by solving the following quadratic programming problem:
\begin{align} \label{svm_sparse}
\min\limits_{\bbeta,\beta_0,\xi_i} &\quad \frac{1}{2} \|\bbeta\|^2+C
\sum_{i=1}^{n}\xi_i\\ 
\text{    s.t.} & \quad \xi_i \geq 0, \quad \forall i, \notag \\
&\quad y_i(\bx_i^{'}\bbeta+\beta_0)\geq1-\xi_i, \quad \forall i, \notag \\
&\quad \| \boldsymbol{\beta} \|_1 \le T. \notag
\end{align}
In the above formulation, the first term is proportional to one over the minimum
distance between a hyperplane that passes from $-\beta_0-1$ and a hyperplane that
passes from $-\beta_0+1$, i.e., one over the thickness of a band (margin) in which we
would like to avoid placing any data points so as to increase the separability
between the two classes. The parameter $C$ is a tunable parameter and $\xi_i$ is a
misclassification penalty for each data point $i$. The constraint on $\|\bbeta\|_1$
imposes sparsity in the variable vector $\bbeta$, thus, allowing only a sparse subset
of features to be selected for the classification decision. The parameter $T$ is also
tunable and controls the level of sparseness.  There exist close connections to
previous work, such as elastic net regularization \cite{zou2005regularization},
$\ell_1$-norm SVM \cite{zhu20041}, and a robust optimization approach for obtaining
appropriate regularizers to learning problems~\cite{chen2018robust}. A drawback of
the formulation (\ref{svm_sparse}) is that it is difficult to kernelize, while
kernelized elastic net has also been proposed in \cite{feng2016kernelized}.

\subsection{JCC: Joint Clustering and Classification} \label{sec:jcc}
In \cite{ACC-2018-PIEEE,xu2016joint} the authors have proposed a Joint Clustering and
Classification (JCC) problem based on the Sparse Linear Support Vector Machine
(SLSVM) framework.  The SLSVM method we saw in Sec.~\ref{sec:slsvm} can in fact be
seen as a special case of JCC where only one cluster is being used. The
classification problem under consideration satisfies the following assumptions. $(i)$
The negative class samples are assumed to be i.i.d. and drawn from a single cluster
with distribution $P_{0}$. $(ii)$ The positive class samples belong to $L$ clusters,
with distributions $P_{1}^{1}, \ldots, P_{1}^{L}$. $(iii)$ Different positive
clusters have different features that separate them from the negative samples (see
Fig.~\ref{figure:asymmetricDemo} for an example).
\begin{figure}[ht]
	\begin{centering}
		\includegraphics[width=0.5\columnwidth]{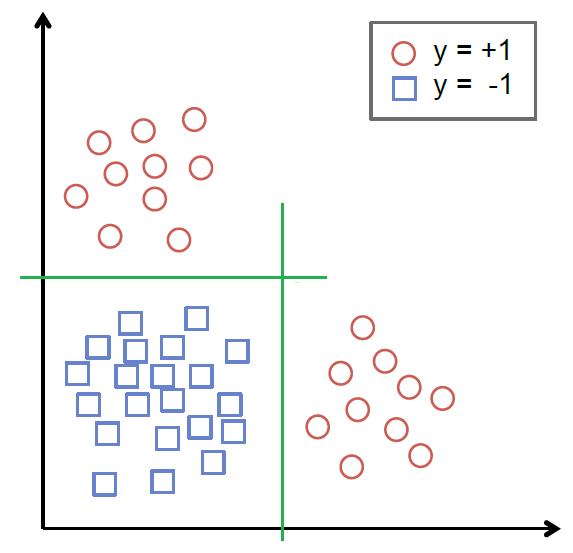}
		\caption{The positive class contains two clusters and each
                 cluster is linearly separable from the negative class.}
		\label{figure:asymmetricDemo}
	\end{centering}
\end{figure}

Let $\mathbf{x}_{i}^{+}$ and $\mathbf{x}_{j}^{-}$ be the $D$ dimensional
positive and negative samples, $y_{i}^{+}, y_{j}^{-}$ the
corresponding labels, where $i\in [N^{+}]$ and $j\in
[N^{-}]$ and $y^{+}_{i} = 1,\ \forall i$ and $y^{-}_{j} =
-1, \ \forall j $.  Assuming $L$ hidden clusters in the positive class,
we try to discover: $(a)$ the $L$ hidden clusters (denoted by a mapping
function $l(i)=l, \ l\in [L]$) and $(b)$ $L$ classifiers
$(\boldsymbol{\beta}^{l}, \beta^{l}_{0})$, as the solution to the
following Joint Clustering and Classification (JCC) problem:
\begin{align}\label{opt:JCC}
\min\limits_{\boldsymbol{\beta}^{l},\beta^{l}_0, l(i)} & \sum\limits_{l=1}^{L} \left(\frac{1}{2}||\boldsymbol{\beta}^{l}||^2 
		+ \lambda^{+}\sum\limits_{i:l(i)=l} \xi_{i}^{l(i)} 
		+ \lambda^{-}\sum\limits_{j=1}^{N^{-}} \zeta_{j}^{l}\right)\\
\text{s.t.} & \sum\limits_{d=1}^D |\beta^{l}_{d}| \leq T^l,\quad
\forall l \in [L], \notag\\
 & \xi_i^{l(i)} \geq 1-y^{+}_i \beta^{l(i)}_0 - \sum\limits_{d=1}^D  y^{+}_i \beta^{l(i)}_{d} x^{+}_{i,d}, 
\; \forall i \in [N^{+}], \notag\\
 & \zeta_j^{l} \geq 1-y^{-}_j \beta^{l}_0 - \sum\limits_{d=1}^D  y^{-}_j
\beta^{l}_{d} x^{-}_{j,d}, 
\; \forall j \in [N^{-}], \forall l \in [L], \notag\\ 
& \xi_i^{l(i)}, \ \zeta_{j}^{l}\geq 0, \quad \forall i \in [N^{+}],
\; \forall j \in [N^{-}], \forall l \in [L], \notag
\end{align}
where $T^l$ is a parameter controlling the sparsity of the classifier in cluster $l$.

In formulation (\ref{opt:JCC}), we have introduced different misclassification
penalties, $\xi_i^{l(i)}$, $\zeta_{j}^{l}$ for positive and negative samples. In
fact, the misclassification costs of the negative samples are counted $L$ times,
since these samples are drawn from a single distribution and are not clustered but
simply copied into each cluster. The parameters $\lambda^{-}$ and $\lambda^{+}$
control the relative weight of these misclassification costs from negative and
positive samples and should be appropriately selected to negate the overcounting,
specifically, we set $\lambda^{+}=L\lambda^{-}$. The constraint $\sum_{d=1}^D
|\beta^{l}_{d}| \leq T^l$ is an $\ell_{1}$-relaxation of the sparsity requirement to
the local classifiers, which is essential to align the formulation with the problem
assumptions and to estimate more robust local classifiers.  The selection of the
tuning parameters is discussed in more detail in \cite{ACC-2018-PIEEE,xu2016joint}.

Two different approaches have been proposed for (\ref{opt:JCC})
\cite{ACC-2018-PIEEE,xu2016joint}. The first, transforms the problem into a Mixed
Integer Programming (MIP) problem but can only solve small-scale problems. The second
approach is an alternating optimization approach which applies to large-scale
problems and also gives rise to theoretical performance guarantees. It is shown in
\cite{ACC-2018-PIEEE,xu2016joint} that it is better to perform joint clustering and
classification instead of separating the two tasks.

\section{SVM based prescriptive analytics} \label{sec:pres}

Prescriptive Support Vector Machines (PSVM) is a prescriptive method we
introduce in this paper that builds on top of SLSVM and JCC.  

Suppose we have generated the per-cluster optimal predictive hyperplanes using
the JCC approach we described in Sec.~\ref{sec:jcc}. Let $\scrC$ be an
index set of variables for each patient we can control/modulate by applying
certain interventions/therapies. For each patient $i$ in the positive class,
with variable vector $\bx_i$, we are interested in optimizing the value of the
controllable variables $x_{i,d}$, for $d\in \scrC$, so that the patient is
predicted to belong to the negative class. 

\begin{figure}[ht]
 	\begin{centering}
 		\includegraphics[width=0.9\columnwidth]{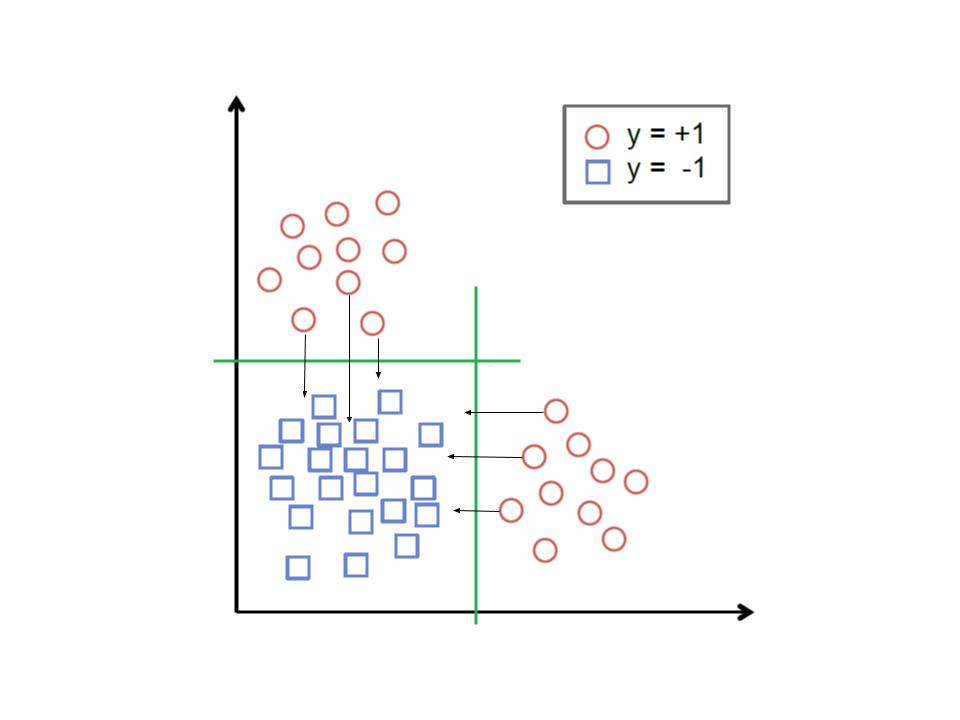}
 		\caption{The readmitted patients moved from the
                  readmitted side of the prediction hyperplane to the
                  non-readmitted side.}
 		\label{figure:prescriptive_Demo}
 	\end{centering}
 \end{figure}
 There is, however, a cost for large changes to the value of the
 controllable variables, which introduces a trade-off between
 ``flipping'' the patient to the negative class and implementing
 interventions that lead to large changes in the controllable variables
 (see Fig.~\ref{figure:prescriptive_Demo}). The following formulation
 optimizes a linear combination of the corresponding two terms in the
 objective. Specifically, consider a patient $i$ in cluster $l$, where
 $i\in [N^{+}]$, $\bx_{i}$ is vector of variables
 characterizing the patient, and $\by_{i}$ is the patient's variables
 after applying the prescription/intervention. Let
 ($\boldsymbol{\beta}^l, \beta_0^l)$ be the coefficients associated with
 the predictive hyperplane discovered by JCC in the $l$-th cluster. To
 determine $\by_{i}$ we solve the following convex optimization problem:
\begin{align} \label{opt:PSVM2}
\min_{\by_i, \xi_i} \quad  & 
\lambda \xi_i + \| \by_{i} -\bx_{i}\|^p_p  \\
\text{   s.t.} \quad  &
\beta_0^l+ (\boldsymbol{\beta}^l)^{'} \by_{i} \leq \xi_i-1, \notag\\ 
\quad &
y_{i,d}=x_{i,d},\; \forall d \not\in \scrC, \notag \\
\quad &
\xi_i \geq 0,\notag \\
\quad &
L_{i,d} \leq y_{i,d} \leq U_{i,d},\; \forall d \in C, \notag 
\end{align}
where $L_{i,d}$ and $U_{i,d}$ are bounds on the controllable variables for each
patient $i$. The parameter $\lambda$ trades-off the failure to flip the patient to
the negative side of the hyperplane with the required change in the patient's
characteristics measured by the term $\| \by_{i} -\bx_{i}\|^p_p$. The higher the
value of $\lambda$, the more attention is given to the goal of preventing a
readmission. To select an appropriate value for $\lambda$ we can use
cross-validation, based on some cost function that accounts for the cost of reducing
readmissions and the cost of prescriptions.

Notice that problem (\ref{opt:PSVM2}) can be solved independently for
each patient who is predicted to belong to the positive class
(readmitted). Thus, it is naturally distributed and can obtain a
prescription for each at-risk patient with only local computations. The
form of the problem (\ref{opt:PSVM2}) depends on the selection of the
$\ell_p$ norm; for instance, when $p=2$, we have a quadratic programming
problem and when $p=1$, we have a linear programming problem.
  
\section{Data and Preprocessing} \label{sec:data}

In this section we describe the data set we use to test and validate our
methods. 

\subsection{NSQIP Dataset Description}

The ACS-NSQIP was created to improve surgical techniques and outcomes
and catalogs over 300 variables on comorbidities, intra-operative events,
and 30-day outcomes using prospective random sampling
\cite{ingraham2010quality}. It contains no protected health information.

The NSQIP dataset contains variables such as:
\begin{itemize}
\item Demographic and health care status characteristics, such as age,
  gender, race, body mass index, smoking, diabetes, hypertension
  requiring medication, and admittance from the emergency room.

\item Procedure information (e.g., CPT code~\cite{cpt}, ICD9 code~\cite{icd9}), the
  American Society of Anesthesiologists (ASA) classification, and wound
  classification.

\item Pre-operative, intra-operative, and post-operative variables,
  including hospital length of stay information, superficial/deep/organ
  space surgical site infections, and existence/description of
  complications (e.g., pneumonia, infections, bleeding, thromboembolic
  events, etc.). 

\item Laboratories, pre-operative and post-operative values.

\end{itemize}

After data pre-processing steps we describe below, there were a total of
$2,288,938$ de-identified patients, $135,293$ of whom were readmitted
within 30 days, resulting in a readmission rate of $5.91\%$.  A total of
230 variables were available for analysis, most of which were binary and
integer with the remaining being continuous.

\subsection{NSQIP Dataset Preprocessing}

Data pre-processing steps we applied were as follows:
\begin{itemize}
\item Patients who died within 30 days from discharge were not included
  in the total of $2,288,938$ patients, as these events compete with
  readmission.

\item Categorical variables (e.g., race, discharge, destination,
  insurance type, CPT code, ICD9 code) were numerically encoded by what
  is typically referred to as {\em one hot encoding}, which amounts to
  introducing a new indicator variable for each category.

\item Missing values of categorical variables were treated as new
  categories and missing values of numerical variables were replaced by
  $k$-nearest-neighbors imputation.

\item Features with small standard deviation ($<0.005$) were removed.

\item One of every two features which were highly linearly correlated
  (absolute value of correlation $>0.8$) was removed.

\item Feature scaling was applied for all features to bring all values
  into the $[0,1]$ range, specifically, all variables were normalized by
  subtracting the minimum and dividing by the range.
\end{itemize}

The variables were further separated into two classes: pre-operative
variables and post-operative variables. Pre-operative variables are
those that can be known before or during the main surgical procedure
while post-operative variables, including complications, can only be
determined after the surgery has been completed.  The reason for
considering these two classes of variables is that some post-operative
variables may be affected by the controllable variables which may be
modulated using our prescriptive method. 

\subsection{Controllable Variables}

We consider three types of controllable variables on which to intervene
using prescriptive analytics:
\begin{itemize}
\item Pre-operative lab tests: sodium, Blood Urea Nitrogen (BUN), serum
  creatinine, serum albumin, bilirubin, SGOT (Serum Glutamic-Oxaloacetic
  Transaminase), alkaline phosphatase, White Blood Cell count (WBC),
  hematocrit (HCT), platelet count, Partial Thromboplastin Time (PTT),
  Prothrombin Time (PT), and International Normalized Ratio (INR) of PT
  values.
\item Length of stay at the hospital: total length of stay, days from
  admission to operation, days from operation to discharge.

\item SSI (Surgical Site Infection) or Infection: occurrences of deep
  incisional SSI, occurrences of organ space SSI, and post-operative
  occurrences of Urinary Tract Infection (UTI).
\end{itemize}

Pre-operative lab values could be altered through appropriate medications and
treatment before the operation to bring them closer to levels not associated with
readmission. The length of stay at the hospital could be to shortened, or lengthened
as appropriate.  Recommendations can also target the tightening of infection control
measures that affect the variables described in the third item above.  In the work we
report in this paper we focus on the pre-operative hematocrit (HCT), as it is a
variable that can be directly impacted (increased) through blood transfusion.  The
predictive models also suggest that pre-operative hematocrit (HCT) is one of the most
important controllable variables.

\section{Performance Evaluation and Experimental Results} \label{sec:num}

\subsection{Prediction Results} 

\subsubsection{Prediction Accuracy} 

In the readmission prediction problem, one typically considers two
distinct performance metrics computed out-of-sample, i.e., over a test
set not seen during training. These metrics are the false positive rate
(or false alarm rate, or one minus the specificity of the test) and the
detection rate (or the true positive rate, or sensitivity of the
test). A Receiver Operating Characteristic (ROC) curve, is a curve that
evaluates the performance of a binary classifier as the decision
threshold is varied, created by plotting the true positive rate against
the false positive rate at different threshold settings. To have a
single metric to compare different ROC curves, we will consider the Area
Under the ROC Curve (AUC). An ideal prediction model has an AUC equal to
$1$, whereas a random prediction would yield an AUC of $0.5$. Anything
with an AUC greater than $0.7$ is considered a moderately good
predictive model.

We randomly chose $60\%$ and $20\%$ of the patients in the dataset to
form the training and validation set and keep the remaining $20\%$ of
the patients as a test set. 

We compared the methods we presented in Sec.~\ref{sec:pred} with some
standard machine learning methods, namely, Random Forest
(RF)~\cite{breiman2001random} and Logistic Regression
(LR)~\cite{friedman2001elements}.  The Random forest
\cite{breiman2001random} is a large collection of decision trees and it
classifies by averaging the decisions of each tree.  Logistic regression
is widely used as a base for comparison in medical machine learning
studies.  In this work, a logistic regression model was fitted with an
additional regularization term: an $\ell_2$-norm term (similar to ridge
regression) \cite{friedman2001elements}.

In Table~\ref{table:NSQIP_AUC}, we compare the performance of the
various classification methods: Random Forests
(RF)~\cite{breiman2001random}, SLSVM, $\ell_2$-regularized logistic
regression (L2LR) with pre-operative variables and post-operative
variables~\cite{friedman2001elements}. JCC was also applied to the
problem but resulted into a single positive cluster, which is identical
to SLSVM. In Table~\ref{table:NSQIP_AUC}, the 2nd column reports AUC
using only pre-operative variables and the 3rd column lists the
corresponding AUC using all (pre-operative and post-operative)
variables.

Methods were implemented in Python (Python Software Foundation,
https://www.python.org/)~\cite{scikit-learn} and Matlab (MathWorks,
Natick, MA). For random forests, the number of trees grown was
500. Cross-validation was used to tune parameters of all methods,
e.g., the number of variables randomly sampled as candidates at each
split for RF, regularization strength for SLSVM and L2LR.
\begin{table}[htbp]
\centering
\caption{Performance of the various classification methods.}
\label{table:NSQIP_AUC}
\begin{tabular}{ccc}
    \hline
    Method & pre-op AUC & post-op AUC  \\  
    \hline
    L2LR & 72.32\% & 83.53\%  \\ 
    RF  & 73.11\% & 84.91\% \\
    SLSVM & 72.28\% & 83.48\% \\   
    \hline
\end{tabular}
\end{table}

Based on the results of Table~\ref{table:NSQIP_AUC}, using
post-operative variables results into substantially better
performance. AUCs of all the methods were similar, perhaps because the
NSQIP dataset contained a large amount of data and a sufficient number of
highly predictive features. It is interesting that we can predict with
such a high accuracy 30-day readmissions. In fact, just this information
can be extremely useful as the health care system can target at-risk
patients and monitor them post-discharge to reduce the risk of
readmission.

\subsubsection{Important Variables} 

For each variable, we computed a two-tailed $p$-value using Welch's $t$-test, where
the null hypothesis was that the two cohorts (readmitted and non-readmitted patients)
have equal means. We found $177$ variables with a $p$-value less than $10^{-6}$.

Using this analysis, the variables with the most statistically
significant values in the two patient cohorts (readmitted and
non-readmitted), were: $(i)$ return to Operating Room (OR) after the
main surgery and before discharge, $(ii)$ length of stay, $(iii)$
occurrences of Surgical Site Infection (SSI, either organ/space SSI,
superficial SSI, deep incisional SSI), $(iv)$ occurrences of urinary
tract infection, $(v)$ occurrences Deep Vein Thrombosis
(DVT)/thrombophlebitis, $(vi)$ occurrences of pulmonary embolism,
$(vii)$ pneumonia occurrences, $(viii)$ estimated probability of
morbidity, $(ix)$ occurrences of sepsis, $(x)$ occurrences myocardial
infarction, $(xi)$ occurrences of progressive renal insufficiency,
$(xii)$ stroke with neurological deficit, $(xiii)$ disseminated cancer,
$(xiv)$ patient currently on dialysis (pre-op), $(xv)$ pre-operative
HCT, $(xvi)$ total operation time (in minutes), and $(xvii)$ Body Mass
Index (BMI).

\subsection{Prescriptive Results}

In this section we evaluate the effectiveness of prescriptions obtained by solving
problem (\ref{opt:PSVM2}) for each patient. We focus on optimizing the patient's
pre-operative hematocrit (HCT) using a blood transfusion.  Transfusions infuse blood
into the patient bloodstream and, typically, a discrete number of bags, each
containing 100cc of blood, gets prescribed. We will limit the number of bags of blood
given to a patient to 3, corresponding to 300cc of blood, which can be considered as
a safe upper limit for blood transfusion. Each bag of blood given to patient
increases HCT by roughly 3\%. We thus define 4 possible treatments as follows:
\begin{itemize}
\item Treatment 1: No transfusion.
\item Treatment 2: 1 bag of blood transfusion.
\item Treatment 3: 2 bags of blood transfusion.
\item Treatment 4: 3 bags of blood transfusion.
\end{itemize}

Since we do not have in the NSQIP data information on whether a blood
transfusion has been performed, we assume a baseline treatment depending
on the patient's HCT as follows:
\begin{itemize}
\item For female patients, if HCT<37, 0 bags of blood are assumed to
  have been given; if 37<HCT<40, 1 bag of blood is assumed to have been
  given; if 40<HCT<43, 2 bags of blood are assumed to have been given;
  and, if HCT>43, 3 bags of blood are assumed to have been given.

\item For male patients, if HCT<41, 0 bags of blood are assumed to have
  been given; if 41<HCT<44, 1 bag of blood is assumed to have been
  given; if 44<HCT<47, 2 bags of blood are assumed to have been given;
  and, if HCT>47, 3 bags of blood are assumed to have been given.
\end{itemize}

We then use formulation (\ref{opt:PSVM2}) to obtain a prescription for
each patient in a test dataset, using the $\ell_2$ norm ($p=2$) in the
penalty associated with the prescribed change in the patient's HCT. In
the absence of ground truth, we then evaluate the effect of the
prescription using a variety of prescriptive methods. We will compare
the readmission rate when prescriptions are being implemented with a
baseline rate set to be equal to the actual readmission rate of $5.85\%$
for patients in the test set.  To be able to compare the readmission
rate with or without the prescriptions, we calibrate each predictive
model by selecting a decision threshold (i.e., a point on the ROC curve
corresponding to the model) so that the model yields the same
readmission rate of $5.85\%$ in the absence of any prescriptions.

Table~\ref{table:NSQIP_rate_updated} reports the results. The first
column lists the predictive model used to evaluate the effects of the
prescriptions. The second column lists the readmission rate after the
optimal prescription is applied to each patient in the test set. The
third column lists the readmission rate in the absence of any
prescriptions.  The average reduction of the readmission rate across the
three predictive models is $1.24\%$, which implies a $21\%$ decrease
compared to the baseline readmission rate of $5.85\%$. The average
percentage change of HCT due to the prescriptions is equal to $4.40\%$.
\begin{table}[htbp]
\centering
\caption{Effect of optimal prescriptions.}
\begin{tabular}{ccc}
  \hline
  Method & Prescriptive rate & Baseline rate \\  
  \hline
  L2LR & 4.95\% & 5.85\%  \\ 
  RF  & 4.70\% & 5.85\% \\
  SLSVM & 4.18\% & 5.85\% \\   
  \hline
\end{tabular}
\label{table:NSQIP_rate_updated}
\end{table}

\section{Conclusions} \label{sec:con}

We developed a new framework to decide prescriptions or other
interventions that reduce the rate of an undesirable event. We build
this prescriptive capability based on an SVM-based predictive
model. Decisions can be decomposed for each subject (patient). 

We applied this new framework to a large dataset of 2.28 million
patients tracked by the ACS-NSQIP over a four year period
(2011--2014). We considered personalized decisions to potentially
increase the pre-operative HCT for each patient through a blood
transfusion. The objective of the prescriptive method is to prevent
30-day readmissions and reduce the corresponding readmission rate.

Our results show that our prescriptive SVM approach reduces the readmission rate by
an average of $1.24\%$ from the readmission rate of $5.85\%$ in the absence of
prescriptions. This amounts to a relative percentage decrease of $21\%$. Considering
that more than \$12 billion were spent in 2005 by Medicare on potentially preventable
readmissions, these types of readmission rate reductions can lead to dramatic savings
on an annual basis.  Future work will consider kernelized methods and speeding them
up them for large-scale datasets.

\section*{Acknowledgments}

We thank Dr. 
George Kasotakis, Dr. 
Dimitris Bertsimas and Michael Lingzhi
Li for useful discussions.  
}

\bibliographystyle{IEEEtran}
\bibliography{psvm}


%
%
\addtolength{\textheight}{-3cm}   

\end{document}

%% file: mymacros_2cln.tex
\newcommand{\be}[1]{\begin{equation}\label{#1}}
\newcommand{\benon}{\begin{equation*}}  
\newcommand{\bemuln}[1]{\begin{multline}\label{#1}}
\newcommand{\bemul}{\begin{multline*}}
\newcommand{\bee}{\begin{eqnarray*}}
\newcommand{\eee}{\end{eqnarray*}}
\newcommand{\been}[1]{\begin{eqnarray}\label{#1}}
\newcommand{\eeen}{\end{eqnarray}}
\newcommand{\began}[1]{\begin{gather}\label{#1}}
\newcommand{\bega}{\begin{gather*}}
\newcommand{\bealn}[1]{\begin{align}\label{#1}}
\newcommand{\beal}{\begin{align*}}
\newcommand{\bealatn}[2]{\begin{alignat}{#1}\label{#2}}
\newcommand{\bealat}{\begin{alignat*}}
\newcommand{\bexalatn}[1]{\begin{xalignat}\label{#1}}
\newcommand{\bexalat}{\begin{xalignat*}}

\newcommand{\mbb}{\mathbb}

{\theoremstyle{plain}

}
{\theoremstyle{break} \theorembodyfont{\it}

 }

%% file: lpsymbols.tex
\def\bx{{\mathbf x}}  
\def\by{{\mathbf y}}

\def\texitem#1{\par\smallskip\noindent\hangindent 25pt
               \hbox to 25pt {\hss #1 ~}\ignorespaces}

\newcommand{\scrC}{\mathcal{C}}

\newcommand{\bbeta}{\boldsymbol{\beta}}